# Observation of Hybrid Magnetic Skyrmion Bubbles in $Fe_3Sn_2$ Nanodisks

Lingyao Kong[1#], Jin Tang[1,2#*], Weiwei Wang[3], Yaodong Wu[4], Jialiang Jiang[2], Yihao Wang[2], Junbo Li[2], Yimin Xiong[1], Mingliang Tian[1,2], and Haifeng Du[2*]

[1]School of Physics and Optoelectronic Engineering, Anhui University, Hefei, 230601, China

[2]Anhui Province Key Laboratory of Condensed Matter Physics at Extreme Conditions, High Magnetic Field Laboratory, HFIPS, Anhui, Chinese Academy of Sciences, Hefei, 230031, China

[3]Institutes of Physical Science and Information Technology, Anhui University, Hefei, 230601,China

[4]School of Physics and Materials Engineering, Hefei Normal University, Hefei, 230601, China

*Corresponding author: jintang@ahu.edu.cn; duhf@hmfl.ac.cn

#These authors contributed equality.

## Abstract

It is well known that there are two types of magnetic bubbles in uniaxial magnets. Here, using Lorentz-transimission electronic microscopy magnetic imaging, we report the direct experimental observation of 3D type-III hybrid bubbles, which comprise Néel-twisted skyrmion bubbles with topological charge $Q = -1$ in near-surface layers and type-II bubbles with $Q = 0$ in interior layers, in Fe$_3$Sn$_2$ nanodisks. Using the tilted magnetic field, we further show the controlled topological magnetic transformations of three types of bubbles in a confined ferromagnetic nanodisk. Our observations are well reproduced using micromagnetic simulations based on measured magnetic parameters. Our results advance fundamental classification and understanding of magnetic bubbles, which could propel the applications of three-dimensional magnetism.

# I. INTRODUCTION

Magnetic bubbles are cylinder domains that have recalled much attention recently because of their similarity to skyrmions[1-9]. The difference is that bubbles are mainly stabilized by magnetic dipole-dipole interaction, while chiral DM interactions stabilize skyrmions [7, 10]. It has been revealed that tiny bubbles with size (~40-90 nm) are comparable with that of skyrmions (~30-70 nm) at room temperature[11, 12]. Furthermore, recent work shows rich bubble dynamics response to electrical stimuli[13], such as current-controlled topological skyrmion-bubble transformations[14], current-induced random helicity reversals of type-I bubbles[15], current-driven dynamic motion of bubbles[16], and current-induced size variations of type-II bubbles[17]. It is thus promising in developing spintronic devices based on traditional magnetic bubbles.

According to the rotation sense of magnetic bubbles, two types of bubbles have been widely established: type-I bubbles with clockwise or counter-clockwise rotations and type-II bubbles with domain-wall orientations lying toward the in-plane field orientations[7-9, 18-20]. The type-I bubble shares the same integer charge $Q$ as chiral skyrmions. Thus, the type-I bubble is also called dipolar skyrmion[12, 16, 20-22], skyrmion bubble[6, 13, 23, 24], bubble skyrmion[25], and topologically-nontrivial bubble[26-28]. Magnetic bubbles are typically stabilized in uniaxial magnets with quality factor $\eta = \frac{2K_u}{\mu_0 M_s^2} > 1$,[7] where $K_u$ is the perpendicular anisotropy, $M_s$ is the saturated magnetization, and $\mu_0$ is the vacuum permeability. Recent studies further show that bubbles could also exist in thick uniaxial magnets with $\eta < 1$.[16, 19] In the case of $\eta < 1$, the competition between perpendicular anisotropy and the dipole-dipole interaction leads to the 3D depth-modulated bubbles, which also explains the complex spin configurations obtained using Lorentz-TEM magnetic imaging.[19] Recent simulations also predict a new type of bubble, which comprises Néel-twisted skyrmion bubbles in near-surface layers and type-II bubbles in interior layers.[18, 29] Such a 3D hybrid bubble state is beyond the traditional definition of bubbles and is named the type-III bubble here. However, direct experimental evidence for stabilizing such type-III bubbles is still lacking.

Here, in combination with 3D micromagnetic simulation based on measured magnetic parameters, we experimentally demonstrate the stabilization of type-III bubbles in confined $Fe_3Sn_2$ nanodisks. The controlled creations of type-III bubbles are also illustrated using tilted magnetic fields. Our results show the diversity of magnetic bubble structures by extended 3D magnetism, which could provide a chance to develop bubble-based spintronics.

## II. METHODS

*Preparation of bulk $Fe_3Sn_2$ crystal*: [19, 30] Single $Fe_3Sn_2$ crystals were grown by chemical vapor transport with stoichiometric iron (Alfa Aesar, >99.9%) and tin (Alfa Aesar, >99.9%). The sintered $Fe_3Sn_2$ was obtained by heating the mixture at 800°C for 7 days, followed by thorough grinding. It was then sealed with $I_2$ in a quartz tube under vacuum and kept in a temperature gradient of 720 °C to 650 °C for 2 weeks.

The $Fe_3Sn_2$ bulk crystal is a lamella, whose out-of-plane orentation is the [001] axis determined by X-ray diffraction .

*Fabrication of $Fe_3Sn_2$ disks*: The 150-nm thick $Fe_3Sn_2$ nanostructured disks with diameters of 510 nm were fabricated from a bulk single crystal using a standard lift-out method, with a focused ion beam and scanning electron microscopy dual beam system (Helios Nanolab 600i, FEI). The ion-sputtered carbon encircled disk was transferred onto a copper grid using a micromanipulator, which was suitable for TEM magnetic imaging. The out-of-plane orientation of the [001] axis of the disk was aligned with that of the bulk $Fe_3Sn_2$ lamella. For detailed information on the nanodisc fabrication process, refer to previous reports [31, 32].

*TEM measurements*: We used in-situ Fresnel imaging in Lorentz-TEM (Talos F200X, FEI) with an acceleration voltage of 200 kV to investigate magnetic domains in the $Fe_3Sn_2$ disk. The DPC microscope was operated at low magnification in scanning TEM (STEM) mode using a split quadrant detector.[19] The probe convergence and detection angles for the DPC-STEM measurements were set to 7 and 1 mrad, respectively; the corresponding probe size is ~3.6 nm. All experiments were performed at room temperature.

*Micromagnetic simulations*: The zero-temperature micromagnetic simulations were performed using MuMax3.[33] We consider the Hamiltonian exchange interaction (*A*) energy, uniaxial magnetic anisotropy ($K_u$) energy, Zeeman energy, dipole-dipole interaction energy. Simulated magnetic parameters are set based on the material

Fe$_3$Sn$_2$ at room temperature ($K_u$ = 54.5 kJ · m$^{-3}$ urated magnetization $M_s$ = 622.7 kA · m$^{-1}$ and $A$ = 8.25 pJ · m$^{-1}$.[19, 30] The simulation cell is set to 3 × 3 × 3 nm$^3$, while the system geometry is defined as a disk with 510 nm diameter and 150 nm thickness.

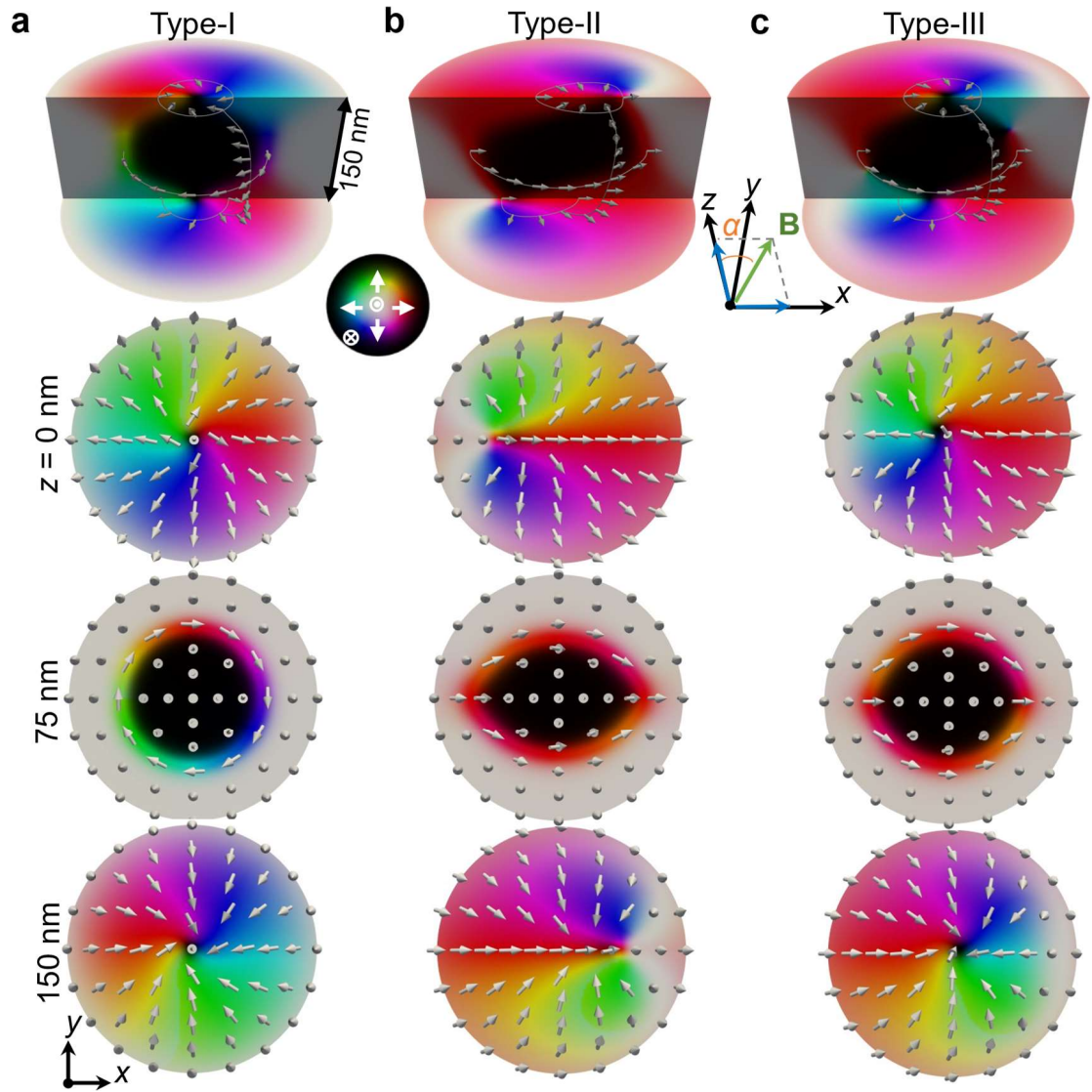

FIG. 1. Simulated 3D magnetic configurations of three types of bubbles at $B$ = 250 mT: a) type-I skyrmion tube with $Q$ = -1 in all layers, b) type-II bubble tube with $Q$ = 0 in all layers, and c) hybrid skyrmion-bubble tube with topological reversals from $Q$

= -1 in the surface layer to $Q$ = 0 in the middle layer. The color represents the magnetization orientation according to the colorwheel. The in-plane component of the magnetic field is along the *x*-axis.

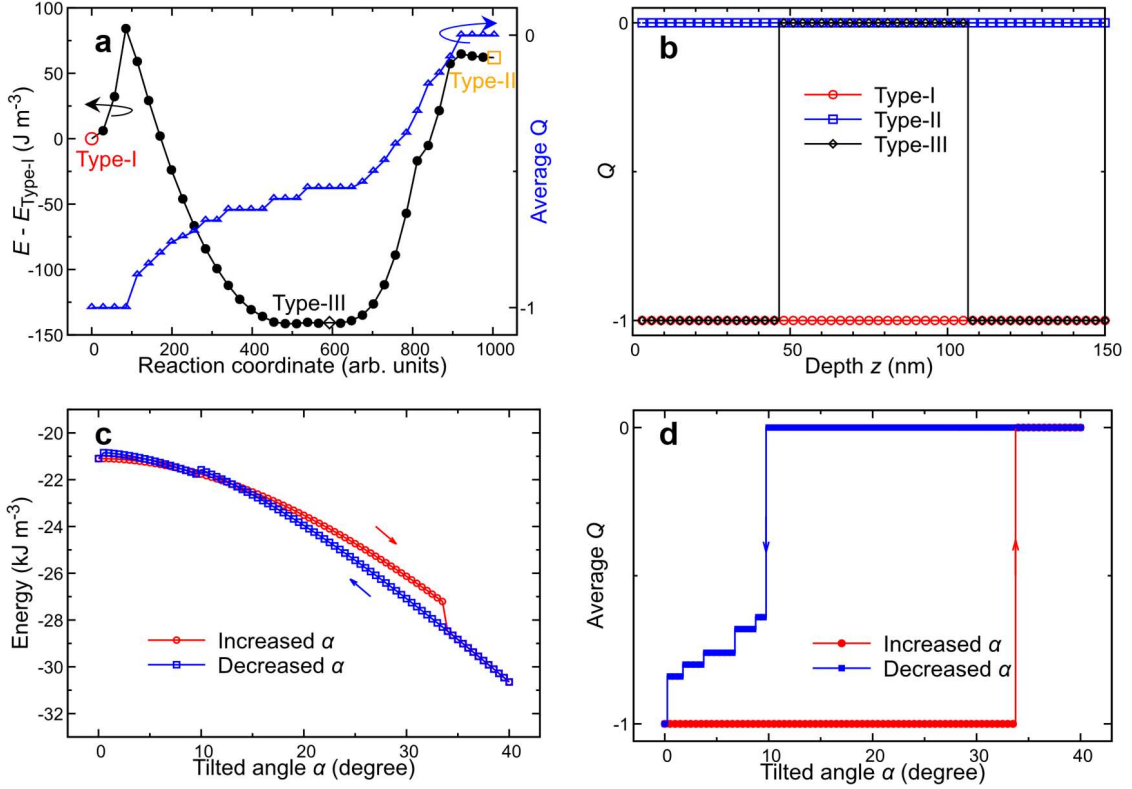

FIG. 2. Theory predicted a type-III bubble in 150-nm thick uniaxial magnet $Fe_3Sn_2$. a) The mutual transformation between type-I and type-II bubbles with the tilted angle of 12° at 0 K. The specified marked symbols corresponding to the stable equilibrium states shown in Figure 1. b) Depth *z* dependence of topological charge for the three types of bubbles. c) and d) Changes of total energy density and average topological charge $Q$ in transformations among the three types of bubbles obtained by tilting the field.

## III. RESULTS AND DISCUSSION

We demonstrate the observation of a type-III bubble in a typical bubble-hosting material $Fe_3Sn_2$.[14, 15, 19, 30, 34, 35] Magnetic bubbles have been discovered in uniaxial ferromagnets for more than 50 years and are categorized into two types[7-9, 18-20]. The type-I skyrmion bubble is typically stabilized by a perpendicular field, while the type-II topologically-trivial bubble will be formed when the field is tilted from the easy axis of uniaxial magnetic anisotropy. Here, we define the tilted angle $\alpha$ as the angle between the magnetic field and the out-of-plane [001] axis of the $Fe_3Sn_2$ nanodisk.

Based on the measured magnetic parameters of $Fe_3Sn_2$, we simulated the 3D magnetic configurations in $Fe_3Sn_2$ nanodisks. The weak quality factor (~0.22) of $Fe_3Sn_2$ contributes to depth-modulated spin twisting of the bubble tube, as shown in Figure 1. The type-I bubble show Néel-twisted skyrmionic configurations in near-surface layers and Bloch-type skyrmion-like structures in interior layers, as shown in Figure 1a. Figure 1b shows a topological trivial type-II bubble under a tilted magnetic field.

At $B$ = 250 mT with a tilted field angle of 12°, our simulation reveals that the two types of bubbles can both stabilize in a 510-nm-diameter $Fe_3Sn_2$ nanodisk. Using the nudged elastic band (NEB) method[36], we obtain the mutual transformation between the two types of bubbles, as shown in Figure 2a. Generally, their mutual transformations at $\alpha = 12°$ must go through a third stable hybrid magnetic phase (Figure 2a and supplemental video 1[37]), a type-III bubble. The type-III bubble shares similar near-surface layers with type-I and similar interior layers with type-II(Figure 1c), so we call the type-III bubble a hybrid bubble. The average topological charges for type-I and type-II bubbles are -1 and 0, respectively (Figure 2b). In contrast, the average topological charge for type-III hybrid bubbles is fractional (Figure 2b). A zero-temperature simulation is conducted due to the fact that the measured room-temperature is significantly smaller than the curie temperature (680 K) of $Fe_3Sn_2$. As an illustration, the energy barrier for bubble transformations such as type-II to type-III bubbles ($\sim 1.08 \times 10^{-19}$ J as shown in Figure 2a) is much larger than the thermal fluctuation energy ($k_B T \sim 4.0 \times 10^{-21}$ J, where $k_B$ is the bolzman constant) at room temperature. This result indicates that the bubbles have excellent thermal stability at room temperature.

The simulated field angle dependence magnetic evolutions at $B$ =250 mT is also studied (Figure 2c and 2d, supplemental Figure S1, and video 2). In the $\alpha$ decreasing process with an initial type-II bubble state at $\alpha = 40°$, a transformation from the type-II bubble to the type-III bubble occurs at $\alpha = 10°$, suggesting that the type-II bubble cannot be stabilized for $\alpha < 10°$. By decreasing the field angle $\alpha$ further, the transformation from the type-III bubble to the type-I bubble happens at $\alpha = 2°$. In contrast, in the $\alpha$ increasing process with an initial type-I bubble state at $\alpha = 0°$, we observe only one topological transformation from type-I to type-II bubbles without the intermediate type-III bubble phase. The simulations suggest that type-III bubbles can be easily achieved in the $\alpha$ decreasing process in a confined ferromagnetic nanodisk. The tilted field angle dependence of bubble transformations is also explained from the view of the energy landscape (Figure 2c and 2d and Supplemental Figure S2). The type-I, II, and III bubbles are the stable phases with the lowest energy at low, high, and intermediate tilted angles, respectively. The energy barrier for the type-I bubble to the type-II or type-III bubble can persist in a wide angular range in the $\alpha$ increasing process (Supplemental Figure S2), suggesting the excellent stability of the type-I bubble in the $\alpha$ increasing process. At a high tilted angle $\alpha = 25°$ at $B = 250$ mT, the tiny energy barrier for the type-I bubble to the type-II bubble also provides the easy transformation from the type-I bubble to the type-II bubble at high tilted field angles. When $\alpha$ decreases to 16°, the type-II bubble raises its energy and becomes unstable. The type-III bubble turns out to be the intermediate phase with the lowest energy between type-I and type-II bubbles. The energy barrier that prevents the collapse of type-II bubbles decreases with the decrement of $\alpha$, resulting in the transformation from the type-II to type-III bubbles. We obtain only type-III and type-I bubble phases in the low tilted field angle. The energy barrier from the type-III bubble to the type-I bubble decreases with the decrement of $\alpha$, providing an easy transformation to the type-I bubble at small tiled field angles.

Theory-predicted mutual transformations among the three types of bubbles are then experimentally explored on the $Fe_3Sn_2$ nanodisk using Lorentz transformation electronic microscopy (TEM). Lorentz-TEM images the in-plane magnetizations with a high spatial resolution[38]. Here, we use both Fresnel Lorentz-TEM and differential phase contrast (DPC) of scanning TEM modes to detect the magnetizations. Figure 3a illustrates that the tilted magnetic field is applied by tilting the sample away from the direction of electron beam irradiation (z-axis). In both experiments and simulations, the in-plane magnetization mappings and corresponding Fresnel images are plotted based on the magnetization component within the xy plane, rather than the disk plane. Moreover, the inclination effect is experimentally identified through the elliptical shape of the effective region in the Fresnel images, as depicted in Figure 3c.

Figure 3b shows the simulated in-plane magnetizations and Fresnel images of three types of bubbles. The overall average in-plane magnetization mapping of the type-II bubble stabilized at $\alpha \sim 20°$ is shown by a uniform domain with magnetizations all pointing to the in-plane field orientation. The average in-plane magnetization of the type-I skyrmion bubble tube is characterized by a two-ring-like vortex, whose internal skyrmion-like core and outer boundary circle spiral are induced by near-surface and interior magnetizations[19], respectively. In contrast, the in-plane magnetic configuration of the type-III hybrid bubble reveals a skyrmion-like core encircled by arch-shaped bubble spirals. The Fresnel contrast of the type-III bubble also reveals hybrid mixed characteristics, including a dot like the type-I bubble and type-II bubble in the boundary. These mixed features of type-III bubbles suggest that we can easily distinguish them from traditional bubbles using Lorentz-TEM. The skyrmion-like configurations in the center and bubble-like configurations in the boundary of type-III bubbles are contributed by magnetizations in near-surface and interior layers (Supplemental Figure S3), respectively.

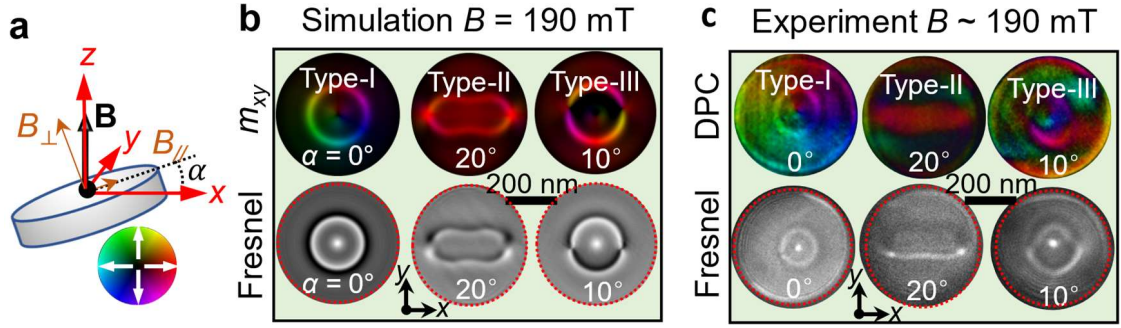

FIG. 3. Experimental observation of the type-III bubble in a Fe$_3$Sn$_2$ nanodisk. a) Sketch map of tilted magnetic field and sample. b) Simulated in-plane magnetization mappings and under-defocused Fresnel images of three types of bubbles at $B = 190$ mT. c) Experimental in-plane magnetization mappings and under-defocused Fresnel images of three types of bubbles at $B \sim 190$ mT. The color represents the in-plane magnetization magnitude and orientation according to the colorwheel. The red dotted circle in (b) are the area of samples. Here, $B_{//} = B\sin\alpha$ denotes the field coponent within the disk plane, and $B_\perp = B\cos\alpha$ represents the field component normal to the disk plane.

Following the theoretical guide in Figures 1 and 2, we experimentally studied the magnetic evolutions induced by tilting field angle in a Fe$_3$Sn$_2$ nanodisk, as shown in Figure 3c, Figure 4, and Figure 5. By increasing the tilted field angle at $B \sim 280$ mT, the type-I bubble at $\alpha \sim 0°$ transforms to the type-II bubble at $\alpha \sim 24°$ without intermediate type-III bubbles (Figure 4a) as predicted in simulations (Figure 2c and 2d). The experimental in-plane magnetization mappings obtained using DPC-STEM, and Lorentz Fresnel images of the type-I and type-II bubbles agree well with our simulations, as shown in Figure 3c. When decreasing the field angle to 0°, we obtain the complex magnetizations (Figure 3c and Figure 4b) in a tilted angular range between 7° and 13° as that of type-III bubbles in simulations (Figure 3b). The in-plane magnetization mappings and Fresnel contrasts of type-III bubbles in our experiments all reveal excellent consistency with that in simulation in a broad field range from ~140 to 320 mT (supplemental Figure S4), which provide the unambiguous experimental proof for the type-III hybrid bubbles. The type-III bubble is always the intermediate magnetic phase from the transformation of the type-II bubble by decreasing the tilted field angle. Because the in-plane component of tilting fields stabilizes the Bloch lines of the type-II bubble domains, the threshold transition angles all decrease as the magnetic field increases, which are also well reproduced in our simulations (Supplemental Figure S1).

The stabilization of type-III bubbles results from the balance between three contributions: the magnetic dipole-dipole interaction in confined nanodisks, the Zeeman energy, and the topological and discrete structural freedom of the magnetic structure. The dipole-dipole interaction primarily contributes to the stabilization of Néel-twisted bubbles with $Q = -1$ in surface layers. At a tilted field condition, the Zeeman energy contributes to the stabilization of bubbles with $Q = 0$ in the interior layers. The discrete structure freedom allows for the topological reversals of type-III bubbles along the depth orientation at an intermediate tilted angle. The dipole-dipole interaction, which can be tuned by thickness, plays an important role in stabilizing type-III bubbles in Fe3Sn2. Our simulations (Supplemental Figure S5) reveal that the in-plane shape anisotropy dominates over the perpendicular anisotropy when the thickness is below 102 nm. Type-III bubbles can be stabilized in a wide range of thicknesses, at least from 102 nm to 510 nm.

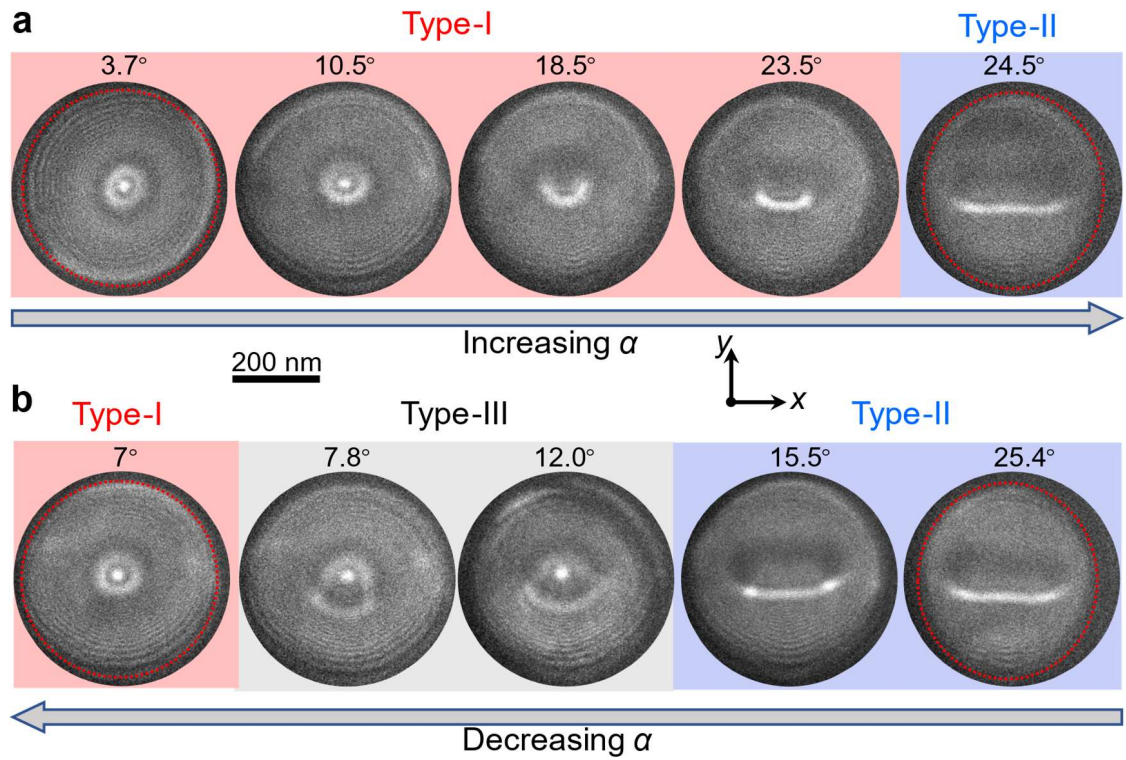

FIG. 4. Bubble transformations at $B \sim 280$ mT with increasing (a) and decreasing (b) of field angle $\alpha$. The red dotted circle are the area of smaples. Defocused distance, -1000 μm.

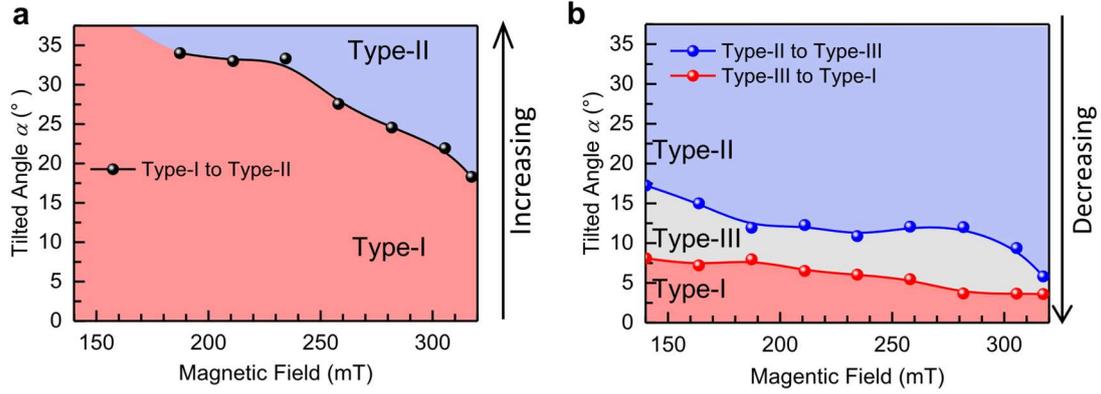

FIG. 5. Experimental field angle dependence of magnetic phase diagram for the three types of bubbles. a) Increasing the field angle from 0° to 35°. Black dots represent the threshold angle for I-to-II bubble transformations. b) Decreasing the field angle from 35° to 0°. Blue and red dots represent the threshold angle for II-to-III and III-to-I bubble transformations, respectively.

## IV. CONCLUSIONS

In summary, in combination with 3D micromagnetic simulations, we have demonstrated the experimental observation of type-III bubbles (i.e., hybrid skyrmion bubbles) with topological reversals along the third dimension of depth. The residual Bloch-skyrmion-like configurations in the center contributed by the near-surface magnetizations can be applied for the experimental distinction of these 3D hybrid skyrmion bubble configurations. Our results should represent an advance in the fundamental classification of magnetic bubbles and the diversity in 3D bubble configurations. Considering the similar topology of the type-I bubble with chiral skyrmions[9, 26], nanoscale bubble size[11, 12], and emergent bubble dynamics to current-induced spin torques[13-17], our results could promote the spintronic devices based on traditional magnetic bubbles.

## ACKNOWLEDGMENTS

This work was supported by the National Key R&D Program of China, Grant No. 2022YFA1403603 and 2021YFA1600200; the Natural Science Foundation of China, Grants No. 12174396, 12104123, and 11974021; Natural Science Project of Colleges and Universities in Anhui Province, Grant No. 2022AH030011; and the Innovation Program for Quantum Science and Technology, Grant No. 2021ZD0302802.

# Supporting information

# Observation of Hybrid Magnetic Skyrmion Bubbles in Fe$_3$Sn$_2$ Nanodisks

Lingyao Kong[1#], Jin Tang[1,2#]*, Weiwei Wang[3], Yaodong Wu[4], Jialiang Jiang[2], Yihao Wang[2], Junbo Li[2], Yimin Xiong[1], Mingliang Tian[1,2], and Haifeng Du[2*]

[1]School of Physics and Optoelectronic Engineering, Anhui University, Hefei, 230601, China

[2]Anhui Province Key Laboratory of Condensed Matter Physics at Extreme Conditions, High Magnetic Field Laboratory, HFIPS, Anhui, Chinese Academy of Sciences, Hefei, 230031, China

[3]Institutes of Physical Science and Information Technology, Anhui University, Hefei, 230601,China

[4]School of Physics and Materials Engineering, Hefei Normal University, Hefei, 230601, China

*Corresponding author: jintang@ahu.edu.cn; duhf@hmfl.ac.cn

[#]These authors contributed equality.

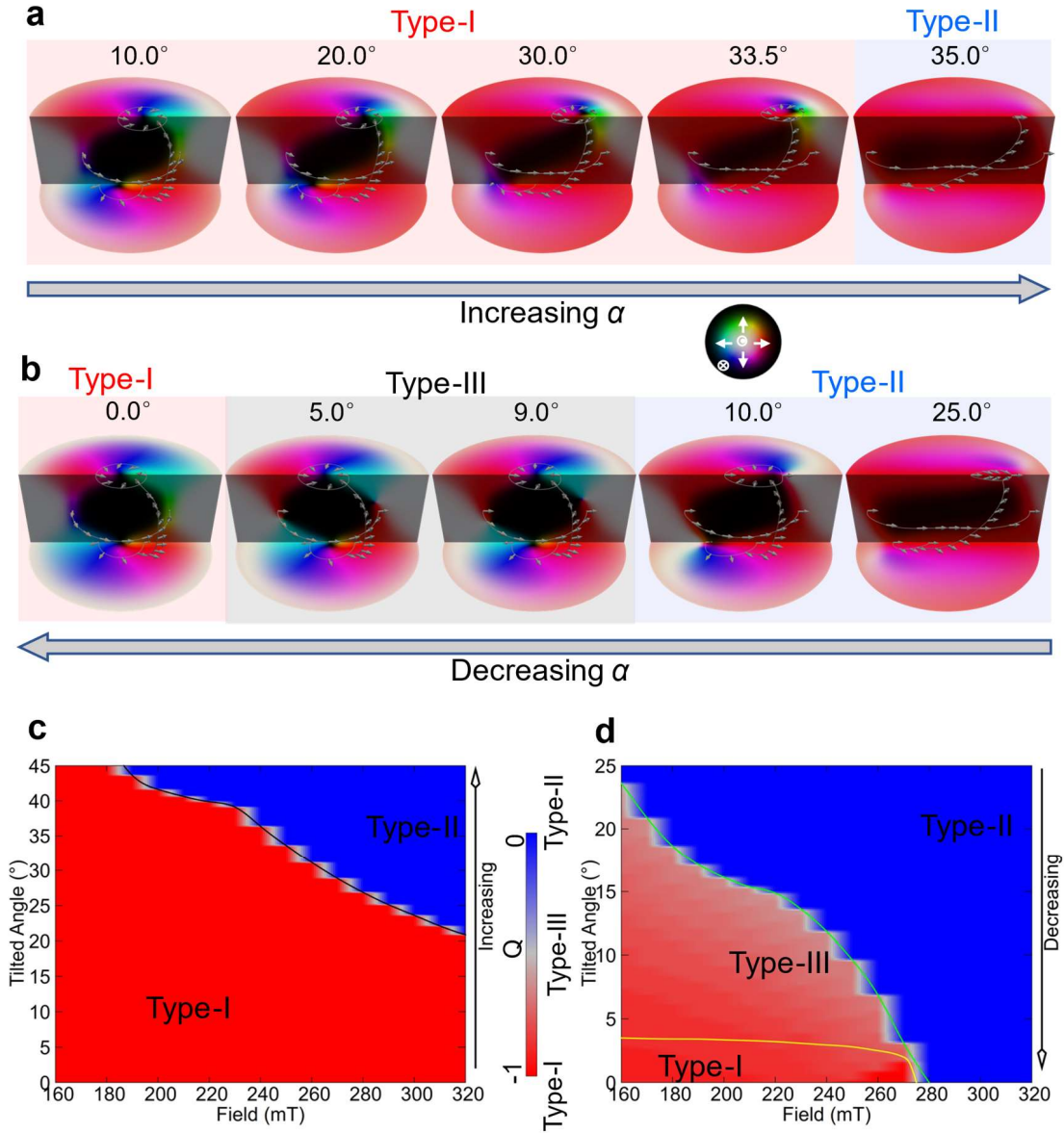

FIG. S1. a) Simulated bubble transformations in the $\alpha$ increasing process at $B = 250$ mT. b) Simulated bubble transformations in the $\alpha$ decreasing process at $B = 250$ mT. c) Simulated magnetic phase diagram for the three types of bubbles in the $\alpha$ increasing process. Black line represents the threshold angle for I-to-II bubble transformations in the $\alpha$ increasing process. (d) Simulated magnetic phase diagram for the three types of bubbles in the $\alpha$ decreasing process. Green and yellow lines represent the threshold angle for II-to-III and III-to-I bubble transformations, respectively, in the $\alpha$ decreasing process. The color in (a) and (b) represents the magnetization orientation according to the colorwheel.

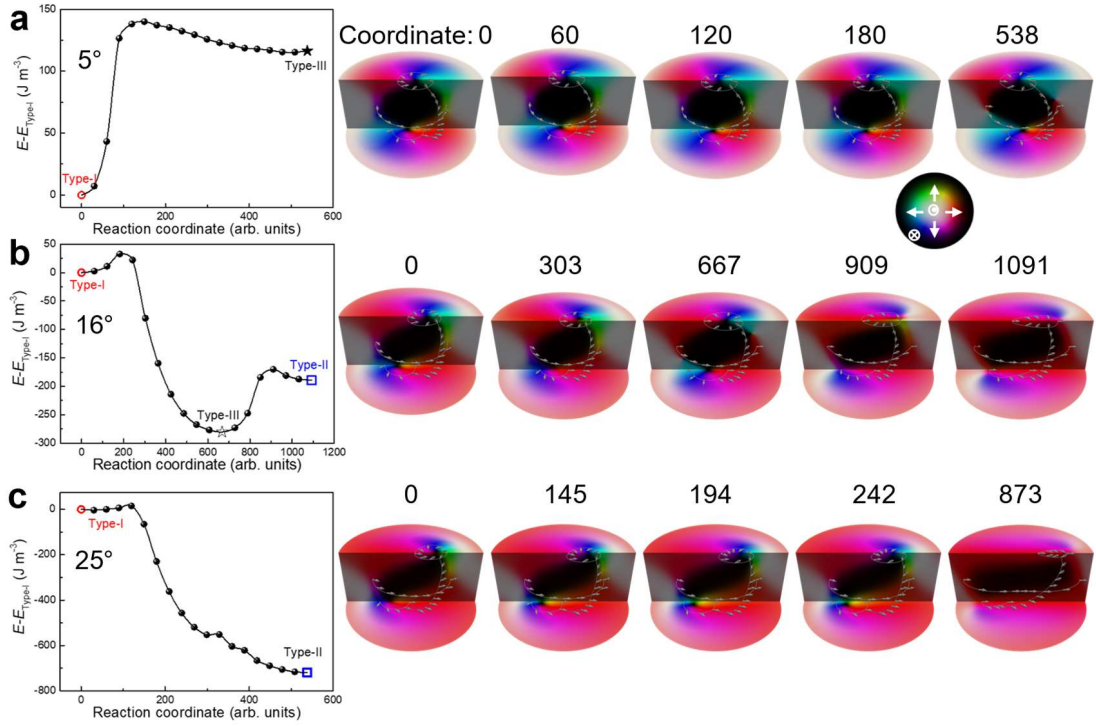

FIG. S2. Simulated bubble transformation at $B = 250$ mT obtained using the NEB method. a) The transformation between type-I and III bubbles at $\alpha = 5°$. b) The transformation between type-I and II bubbles at $\alpha = 16°$. c) The transformation between type-I and II bubbles at $\alpha = 25°$. The color represents the magnetization orientation according to the colorwheel.

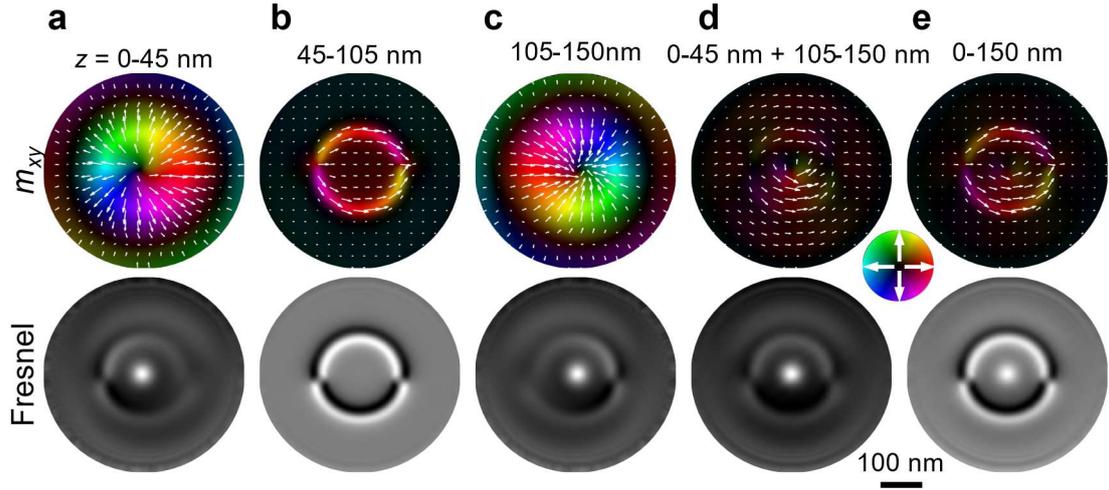

FIG. S3. Simulated in-plane magnetization mappings and corresponding over-defocused Fresnel images of the type-III bubble textures in Fe$_3$Sn$_2$ for magnetizations in a) the bottom surface layers $z = 0 - 45$ nm, b) the interior layers $z = 45 - 105$ nm, c) the top surface layers $z = 105$-$150$ nm, d) the near-surface layers $z = 0 - 45$ nm and 105-150 nm, and e) overall layers $z = 0$-$150$ nm. The color in (a) represents the in-plane magnetization orientation and magnitude according to the colorwheel.

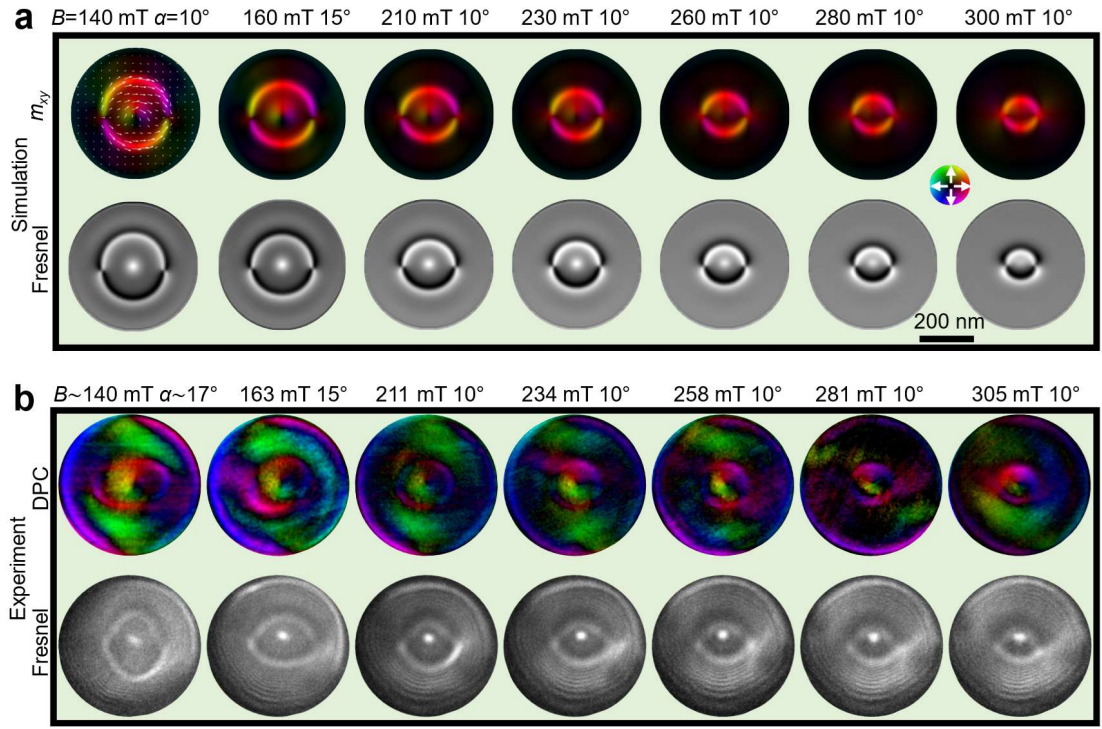

FIG. S4. a) Simulated and b) experimental magnetic field dependence of evolutions of the type-III bubble in the $Fe_3Sn_2$ nanodisk. The color in (a) and (b) represents the in-plane magnetization orientation and magnitude according to the colorwheel in (a).

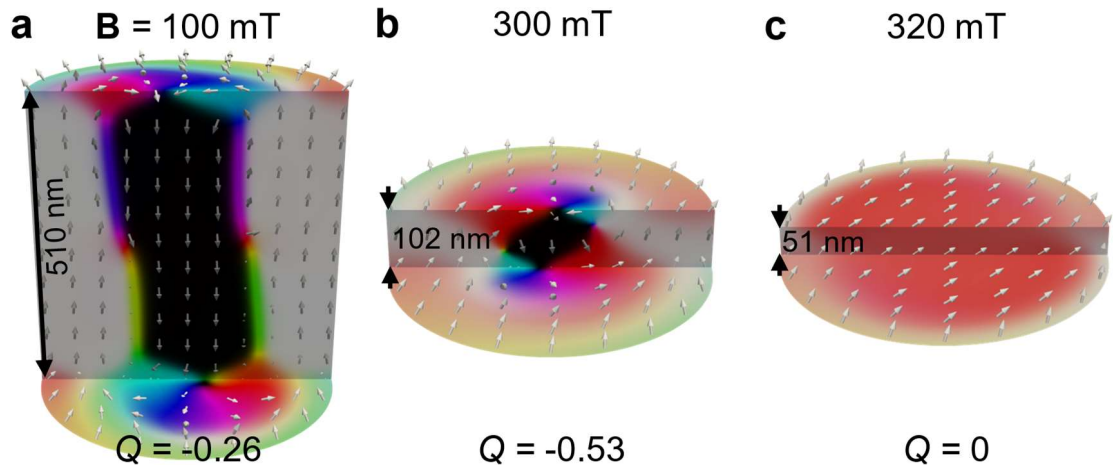

FIG. S5. Simulated thickness dependence stability of type-III bubble in the Fe$_3$Sn$_2$ nanodisk, for a), thickness 510 nm and magnetic field 100 mT, b), thickness 102 nm and magnetic field 300 mT, c). thickness 51 nm and magnetic field 320 mT. The tilted angle of magnetic field 10°.

**Supplementary Video Captions:**

**Movie S1.** Simulated transformation between the type-I and type-II bubbles at $B = 250$ mT and $\alpha = 12°$ obtained using the NEB method.

**Movie S2.** Simulated magnetic evolution in the $\alpha$ decreasing and increasing process at $B = 250$ mT.